# Thermodynamics of effective Minkowski spacetime in self-assembled hyperbolic metamaterials

Igor I. Smolyaninov

*Department of Electrical and Computer Engineering, University of Maryland, College Park, MD 20742, USA*

**Recent developments in gravitation theory indicate that the classic general relativity is an effective macroscopic theory which will be eventually replaced with a more fundamental theory based on thermodynamics of yet unknown microscopic degrees of freedom. Here we consider thermodynamics of an effective Minkowski spacetime which may be formed under the influence of external magnetic field in a cobalt ferrofluid. It appears that the extraordinary photons propagating inside the ferrofluid perceive thermal gradients in the ferrofluid as an effective gravitational field, which obeys the Newton law. Moreover, the effective Minkowski spacetime behaviour near the metric signature transition may mimic various cosmological Big Bang scenarios, which may be visualized directly using an optical microscope. Thus, some important features of the hypothetic microscopic theory of gravity are reproduced in the ferrofluid-based analogue model.**

Modern advances in gravitation theory [1,2] strongly indicate that the classic general relativity description of gravity results from a more fundamental theory based on thermodynamics of yet unknown microscopic degrees of freedom.    These true microscopic degrees of freedom cannot be probed directly because they are far out of reach of terrestrial experiments and astronomical observations. Therefore, it is instructive to look at various examples of emergent gravity in various analogue spacetimes, which appear in several condensed matter systems [3]. By probing such systems we may gain valuable insights into the relationship between the macroscopic effective gravity and the thermodynamics of well-known microscopic atomic degrees of freedom.

Ferrofluid-based self-assembled hyperbolic metamaterials have emerged recently as an interesting example of a condensed matter system, which may be described by an emergent effective Minkowski spacetime, and which exhibits gravity-like nonlinear optical interactions [4-6]. If a ferrofluid is subjected to a strong enough external magnetic field, it self-assembles into a "wire array" hyperbolic metamaterial (see Fig.1), which may exhibit reach physics associated with topological defects [8] and metric signature transitions [9]. Moreover, such a self-assembly may occur even without the external field at low enough temperature, due to spontaneous magnetization of the ferrofluid. As was pointed out recently by Mielczarek and Bojowald [10,11], such self-assembled magnetic nanoparticle-based hyperbolic metamaterials exhibit strong similarities with the properties of some microscopic quantum gravity models, such as loop quantum cosmology.

Here we will consider thermodynamics of the effective Minkowski spacetime in a ferrofluid. It appears that the extraordinary photons propagating inside the ferrofluid perceive thermal gradients in the ferrofluid as an effective gravitational field, which obeys the Newton law. Moreover, the effective Minkowski spacetime behaviour near the metric signature transition may mimic various cosmological Big Bang scenarios. Thus, some important features of the hypothetic microscopic theory of gravity are reproduced in the ferrofluid-based analogue model. Even though the relationship between the thermodynamics and the effective gravity in 3D ferrofluid spacetime does not completely match the behaviour of general relativity in four spacetime dimensions, this model system is quite interesting. It permits both direct theoretical analysis at the level of well-defined microscopic degrees of freedom, and it enables direct visualization of these microscopic degrees of freedom using standard optical tools.

As a starting point, let us consider the electromagnetic properties of self-assembled ferrofluids based on their microscopic degrees of freedom. As an example, let us choose the cobalt magnetic fluid 27-0001 from Strem Chemicals used in experimental studies reported in [4,7,8], which is composed of 10 nm magnetic cobalt nanoparticles in kerosene. The nanoparticles are coated with sodium dioctylsulfosuccinate and a monolayer of LP4 fatty acid condensation polymer. The average volume fraction of cobalt nanoparticles in this ferrofluid is p=8.2%. Magnetic nanoparticles in this ferrofluid are known to form nanocolumns aligned along the magnetic field [7]. Macroscopic electromagnetic properties of such a ferrofluid may be understood based on the Maxwell-Garnett approximation via the dielectric permittivities $\varepsilon_m$ and $\varepsilon_d$ of cobalt and kerosene, respectively. Volume fraction of cobalt nanoparticles aligned into nanocolumns by the external magnetic field, $\alpha(H,T)$, depends on temperature and the field magnitude. At very large magnetic fields all nanoparticles are aligned into nanocolumns, so that $\alpha(\infty,T) = $p=8.2%, and a 3D wire array hyperbolic metamaterial [4,5] is formed as shown schematically in Fig.1(a). We will assume that at smaller fields the difference $\alpha(\infty,T) - \alpha(H,T)$ describes cobalt nanoparticles, which are not aligned and distributed homogeneously inside the ferrofluid. Dielectric polarizability of these nanoparticles may be included into $\varepsilon_d$, leading to slight increase in its value. Using this model, the diagonal components of the ferrofluid permittivity may be calculated using the Maxwell-Garnett approximation as follows [12]:

$$\varepsilon_z = \varepsilon_2 = \alpha(H,T)\varepsilon_m + (1 - \alpha(H,T))\varepsilon_d \qquad (1)$$

$$\varepsilon_x = \varepsilon_y = \varepsilon_1 = \frac{2\alpha(H,T)\varepsilon_m\varepsilon_d + (1 - \alpha(H,T))\varepsilon_d(\varepsilon_d + \varepsilon_m)}{(1 - \alpha(H,T))(\varepsilon_d + \varepsilon_m) + 2\alpha(H,T)\varepsilon_d} \qquad (2)$$

Calculated wavelength dependencies of $\varepsilon_2$ and $\varepsilon_1$ at $\alpha(\infty,T) = p = 8.2\%$ are plotted in Fig. 1(b). These calculations are based on the optical properties of cobalt in the infrared range [13]. While $\varepsilon_1$ stays positive and almost constant, $\varepsilon_2$ changes sign to negative around $\lambda = 1\mu m$. If the volume fraction of cobalt nanoparticles varies, this change of sign occurs at some critical value $\alpha_H$:

$$\alpha > \alpha_H = \frac{\varepsilon_d}{\varepsilon_d - \varepsilon_m} \quad , \qquad (3)$$

so that the ferrofluid becomes a hyperbolic metamaterial [4,5]. Alignment of cobalt nanoparticle chains along the direction of external magnetic field is clearly revealed by microscopic images of the ferrofluid, as shown in Fig.1(c). As far as magnetic permeability is concerned, at the visible and infrared frequencies the ferrofluid may be considered as a non-magnetic (μ=1) medium.

As a second step, let us demonstrate that the wave equation describing propagation of monochromatic extraordinary light inside a hyperbolic metamaterial exhibits 2+1 dimensional Lorentz symmetry. A detailed derivation of this result may be found in [5]. Since the ferrofluid in external magnetic field may be characterized as a uniaxial dielectric material, the electromagnetic field inside the ferrofluid may be separated into ordinary and extraordinary waves. As a reminder, vector $\vec{E}$ of the extraordinary light wave is parallel to the plane defined by the $k$–vector of the wave and the optical axis of the metamaterial. Propagation of monochromatic extraordinary light may be described by a coordinate-dependent wave function $\varphi_\omega = E_z$ obeying the wave equation:

$$-\frac{\omega^2}{c^2}\varphi_\omega = \frac{\partial^2\varphi_\omega}{\varepsilon_1\partial z^2} + \frac{1}{\varepsilon_2}\left(\frac{\partial^2\varphi_\omega}{\partial x^2} + \frac{\partial^2\varphi_\omega}{\partial y^2}\right) \quad (4)$$

This wave equation coincides with the Klein-Gordon equation for a massive scalar field $\varphi_\omega$ in 3D Minkowski spacetime:

$$-\frac{\partial^2\varphi_\omega}{\varepsilon_1\partial z^2} + \frac{1}{(-\varepsilon_2)}\left(\frac{\partial^2\varphi_\omega}{\partial x^2} + \frac{\partial^2\varphi_\omega}{\partial y^2}\right) = \frac{\omega_0^2}{c^2}\varphi_\omega = \frac{m^{*2}c^2}{\hbar^2}\varphi_\omega \quad (5)$$

in which the spatial coordinate $z$ behaves as a "timelike" variable. Eq.(5) exhibits effective Lorentz invariance under the coordinate transformation

$$z' = \frac{1}{\sqrt{1 - \frac{\varepsilon_{xy}}{(-\varepsilon_z)}\beta}}(z - \beta x) \quad (6)$$

$$x' = \frac{1}{\sqrt{1 - \frac{\varepsilon_{xy}}{(-\varepsilon_z)}\beta}}\left(x - \beta\frac{\varepsilon_{xy}}{(-\varepsilon_z)}z\right),$$

where $\beta$ is the effective boost. Similar to our own Minkowski spacetime, the effective Lorentz transformations in the $xz$ and $yz$ planes form the Poincare group together with translations along $x$, $y$, and $z$ axis, and rotations in the $xy$ plane. Thus, wave equation (5) describes world lines of massive particles (with an effective mass $m^* = \hbar\omega_0/c^2$), which propagate in an effective 2+1 dimensional Minkowski spacetime. Note that the components of the metamaterial dielectric tensor define the effective metric $g_{ik}$ of this spacetime: $g_{00} = -\varepsilon_1$ and $g_{11} = g_{22} = -\varepsilon_2$.

When the nonlinear optical effects become important, they are described in terms of various order nonlinear susceptibilities $\chi^{(n)}$ of the metamaterial:

$$D_i = \chi^{(1)}_{ij} E_j + \chi^{(2)}_{ijl} E_j E_l + \chi^{(3)}_{ijlm} E_j E_l E_m + ... \qquad (7)$$

Taking into account these nonlinear terms, the dielectric tensor of the metamaterial (which defines its effective metric) may be written as

$$\varepsilon_{ij} = \chi^{(1)}_{ij} + \chi^{(2)}_{ijl} E_l + \chi^{(3)}_{ijlm} E_l E_m + ... \qquad (8)$$

In a centrosymmetric material all the second order nonlinear susceptibilities $\chi^{(2)}_{ijl}$ must be equal to zero. It is clear that eq.(8) provides coupling between the matter content (photons) and the effective metric of the metamaterial spacetime. Nonlinear optical effects "bend" this effective Minkowski spacetime, resulting in gravity-like interaction of extraordinary light rays. It appears that in the weak field limit the nonlinear optics of hyperbolic metamaterials may indeed be formulated as an effective gravity [5]. In such a limit the Einstein equation

$$R_i^k = \frac{8\pi\gamma}{c^4} \left( T_i^k - \frac{1}{2} \delta_i^k T \right) \qquad (9)$$

is reduced to

$$R_{00} = \frac{1}{c^2} \Delta\phi = \frac{1}{2} \Delta g_{00} = \frac{8\pi\gamma}{c^4} T_{00} \quad , \qquad (10)$$

where $\phi$ is the gravitational potential [14]. Therefore, the third order terms in eq.(8) may provide correct coupling between the effective metric and the energy-momentum tensor. These terms are associated with the optical Kerr effect. The detailed analysis in [5] indeed indicates that the Kerr effect in a hyperbolic metamaterial leads to effective gravity.

Let us briefly reproduce this analysis, since it is important for the discussion of connection between the effective gravity and the thermodynamics of a magnetized ferrofluid. Since $z$ coordinate plays the role of time, while $g_{00}$ is identified with $-\varepsilon_1$, eq.(10) may be re-written as

$$-\Delta^{(2)}\varepsilon_1 = \frac{16\pi\gamma*}{c^4}T_{zz} = \frac{16\pi\gamma*}{c^4}\sigma_{zz} \quad , \tag{11}$$

where $\Delta^{(2)}$ is the 2D Laplacian operating in the $xy$ plane, $\gamma*$ is the effective "gravitation constant", and $\sigma_{zz}$ is the $zz$ component of the Maxwell stress tensor of the electromagnetic field in the medium:

$$\sigma_{zz} = \frac{1}{4\pi}\left(D_z E_z + H_z B_z - \frac{1}{2}\left(\vec{D}\vec{E} + \vec{H}\vec{B}\right)\right) \tag{12}$$

Taking into account eq.(4), for a single plane wave eq.(11) may be rewritten as [5]

$$-\Delta^{(2)}\varepsilon_1 = -\Delta^{(2)}\left(\varepsilon_1^{(0)} + \delta\varepsilon_1\right) = k_x^2\delta\varepsilon_1 = -\frac{4\gamma*B^2 k_z^2}{c^2\omega^2\varepsilon_1} \quad , \tag{13}$$

where without a loss of generality we have assumed that the $B$ field of the wave is oriented along $y$ direction. We also assumed that the nonlinear corrections to $\varepsilon_1$ are small, so that we can separate $\varepsilon_1$ into the constant background value $\varepsilon_1^{(0)}$ and the weak nonlinear corrections. These nonlinear corrections do indeed look like the Kerr effect assuming that the extraordinary photon wave vector components are large compared to $\omega/c$:

$$\delta\varepsilon_1 = -\frac{4\gamma*B^2 k_z^2}{c^2\omega^2\varepsilon_1 k_x^2} \approx \frac{4\gamma*B^2}{c^2\omega^2\varepsilon_2} = \chi^{(3)}B^2 \tag{14}$$

This assumption has to be the case if extraordinary photons may be considered as classic "particles". Unlike the usual "elliptic" optical materials, this assumption is justified by the hyperbolic dispersion relation of the extraordinary photons:

$$\frac{\omega^2}{c^2} = \frac{k_z^2}{\varepsilon_1} - \frac{k_x^2 + k_y^2}{|\varepsilon_2|}$$

(15)

which follows from eq.(4). Eq.(14) establishes connection between the effective gravitation constant $\gamma^*$ and the third order nonlinear susceptibility $\chi^{(3)}$ of the hyperbolic metamaterial. Since $\varepsilon_2 < 0$, the sign of $\chi^{(3)}$ must be negative for the effective gravity to be attractive. This condition is satisfied naturally in most liquids, and in particular, in kerosene. Because of the large and negative thermo-optic coefficient inherent to most liquids, heating produced by partial absorption of the propagating beam translates into a significant decrease of the refractive index at higher light intensity. For example, the reported thermo-optic coefficient of water reaches $\Delta n/\Delta T = -5.7 \times 10^{-4} \text{K}^{-1}$ [15]. Moreover, introduction of nanoparticles or absorbent dye into the liquid allows for further increase of the thermal nonlinear response [16]. Therefore, a ferrofluid-based self-assembled hyperbolic metamaterial naturally exhibits effective gravity. The thermal origin of this effective gravity looks interesting in light of the modern advances in gravitation theory [1,2], which strongly indicate that the classic general relativity description of gravity results from thermodynamic effects.

As a next step, let us consider the effect of thermal gradients in the ferrofluid on its effective metric. It appears that the extraordinary photons propagating inside the ferrofluid perceive thermal gradients as an effective gravitational field. Indeed, eqs. (1,2) imply that $\varepsilon_2$ and $\varepsilon_1$ (which may be understood as the effective metric coefficients $g_{00} = -\varepsilon_1$ and $g_{11} = g_{22} = -\varepsilon_2$ of the metamaterial spacetime) depend on the volume fraction $\alpha(H,T)$ of cobalt nanoparticles aligned into nanocolumns by the external magnetic field, which in turn depends on the local temperature of the ferrofluid. Since ferrofluids subjected to external magnetic field are known to exhibit classical superparamagnetic behaviour [17], well established results from the theory of magnetism may be used to calculate $\alpha(H,T)$. Superparamagnetism is a form of magnetism that is exhibited by magnetic materials, which consist of small ferromagnetic or ferrimagnetic nanoparticles. Superparamagnetism occurs in nanoparticles which are single-domain, which is possible when their diameter is below ~50 nm, depending on the material. Since the typical size of magnetic nanoparticles in ferrofluids is ~ 10 nm, magnetic ferrofluids also belong to the class of superparamagnetic materials. When an external magnetic field is applied to an assembly of superparamagnetic nanoparticles, their magnetic moments tend to align along the applied field, leading to a net magnetization. If all the particles may be considered to be roughly identical (as in the case of a homogeneous ferrofluid), and the temperature is low enough, then the magnetization of the assembly is [17]

$$M(H,T) = n\mu \tanh\left(\frac{\mu H}{kT}\right) \quad , \tag{16}$$

where $n$ is the nanoparticle concentration, and $\mu$ is their magnetic moment. Therefore, within the scope of our model of the dielectric response of the ferrofluid (see eqs.(1,2)), we may assume that

$$\alpha(H,T) = \alpha_\infty \tanh\left(\frac{\mu H}{kT}\right) \tag{17}$$

where $\alpha_\infty = 0.082$. Let us consider the typical case of $-\varepsilon_m >> \varepsilon_d$ and assume that $\alpha(H,T)$ is small (which is typically required for the Maxwell-Garnett approximation to be valid). In such a case the effective metric coefficients are:

$$g_{11} = g_{22} = -\varepsilon_2 \approx -\alpha_\infty \tanh\left(\frac{\mu H}{kT}\right)\varepsilon_m - \varepsilon_d \tag{18}$$

$$g_{00} = -\varepsilon_1 \approx -\varepsilon_d \left( 1 + 2\alpha_\infty \tanh\left( \frac{\mu H}{kT} \right) \right) \qquad (19)$$

The effective spacetime appears to be a Minkowski one if the temperature is low enough, so that

$$\alpha_\infty \tanh\left( \frac{\mu H}{kT} \right) > \frac{\varepsilon_d}{\left( -\varepsilon_m \right)} \qquad (20)$$

Thus, in the weak field limit the effective gravitational potential (see eq.(10)) is

$$\phi = \frac{c^2}{2} \left( \frac{g_{00}}{\varepsilon_d} - 1 \right) = \alpha_\infty c^2 \left( \tanh\left( \frac{\mu H}{kT} \right) - 1 \right) \qquad (21)$$

(where the value of the effective potential at $T$=0 is chosen as a reference) and the effective gravitational field $\vec{G}$ is

$$\vec{G} = -\nabla \phi = \frac{\alpha_\infty c^2 \mu H}{kT^2 \cosh^2\left( \frac{\mu H}{kT} \right)} \nabla T \qquad (22)$$

where we have assumed the external magnetic field $H$ to be constant. Thus, the extraordinary photons propagating inside the ferrofluid perceive thermal gradients in the ferrofluid as an effective gravitational field. This observation correlates nicely with the thermal origin of gravity-like nonlinear optical interaction of the extraordinary light rays due to photo-thermal Kerr effect, which was noted above.

Let us consider a linear source of heat $q$ (see Fig.2) placed inside the ferrofluid, which is constant in time and aligned parallel to the external magnetic field $H$ (and therefore it is also parallel to the nanoparticle chains). Let us also assume that the ferrofluid is kept at a constant temperature $T_0$. The temperature distribution around the source is defined by the two-dimensional heat conductance equation

$$-\sigma\left(\frac{\partial^2}{\partial x^2}+\frac{\partial^2}{\partial y^2}\right)T=-\sigma\nabla(\nabla T)=\frac{1}{c_p\rho}q \tag{23}$$

where $\sigma$ is the thermal diffusivity, $c_p$ is the heat capacity and $\rho$ is the density of the ferrofluid. If the linear source of heat is weak enough, so that the temperature-dependent terms in the denominator of eq.(22) may be considered constant, eq.(23) may be re-written as an equation for the effective gravitational field $\vec{G}$ :

$$\nabla\vec{G}=-\frac{\alpha_\infty c^2\mu H}{\sigma c_p\rho kT_0^2\cosh^2\left(\dfrac{\mu H}{kT_0}\right)}q=-\gamma*q \tag{24}$$

which has the form of the Newton law of gravity. As evident from eq.(24), the heat source $q$ plays the role of a gravitational mass. Thus, a linear source of heat parallel to the external magnetic field $H$ behaves as a world line of massive gravitating object in the metamaterial Minkowski spacetime. It is interesting that unlike quantum mechanical derivation based on the holographic principle reported in [2], the Newton law in a ferrofluid arises in a purely classical Boltzmann system. In principle, a similar approach may be applied to other magnetic systems at lower temperatures, where the thermodynamics description becomes explicitly quantum mechanical [19]. However, direct visualization of gravity-like effects in such systems appears to be much more difficult.

As far as the strong field limit is concerned (where the temperature may not be considered almost constant) it was pointed out in ref.[5] that at large enough power gravitational self-interaction of the extraordinary rays is strong enough, so that spatial solitons may be formed. These spatial solitons also behave as world lines of compact self-gravitating bodies in the effective 2+1 dimensional Minkowski spacetime. This is not surprising since due to photo-thermal effect solitons may also be considered as linear sources of heat. Moreover, upon increase of the optical power, a spatial soliton may collapse into a black hole analogue [5]. Indeed, in the presence of self-defocusing negative Kerr effect in the dielectric host $\varepsilon_d$, the wave equation (eq.(5)) must be modified. Assuming a spatial soliton-like solution which conserves energy per unit length $W \sim P/c$ (where $P$ is the laser power), the soliton width $\rho$ and the magnetic field amplitude $B$ of the extraordinary wave are related as

$$B^2 \rho^2 = P/c \qquad (25)$$

As a result, eq.(5) must be re-written as

$$-\frac{\partial^2 \varphi_\omega}{\left(\varepsilon_1^{(0)} - \frac{\left(-\chi^{(3)}\right)P}{c\rho^2}\right)\partial z^2} + \frac{1}{\left(-\varepsilon_2\right)}\left(\frac{\partial^2 \varphi_\omega}{\partial x^2} + \frac{\partial^2 \varphi_\omega}{\partial y^2}\right) = \frac{\omega_0^2}{c^2}\varphi_\omega \qquad (26)$$

where $\varepsilon_1^{(0)}$ is the dielectric permittivity component at $P$=0 (note that the nonlinear contribution to $\varepsilon_2 \approx \alpha\varepsilon_m$ may be neglected). Effective metric described by eq.(26) has a black hole-like singularity at

$$\rho = \left(\frac{\left(-\chi^{(3)}\right)P}{c\varepsilon_1^{(0)}}\right)^{1/2} \qquad (27)$$

The critical soliton radius at $P$=100W was estimated to be ~20 nm [5], which does not look completely unrealistic from the fabrication standpoint. While achieving such critical values with CW lasers seems implausible, using pulsed laser definitely looks like a realistic option since thermal damage produced by a pulsed laser typically depends on the pulse energy and not the pulse power.

Another "strong field" situation corresponds to the ferrofluid behavior near the metric signature transition. It appears that "melting" of the effective Minkowski spacetime in such a case may mimic various cosmological Big Bang scenarios. This is not surprising since symmetry breaking in magnetic systems is typically described by the Mexican hat potential. Moreover, such a Big Bang-like behavior may be observed directly using an optical microscope. Indeed, based on eq.(5) and Fig.1(b) it is clear that the factor *(-ε₂)* plays the role of a scale factor of the effective Minkowski spacetime

$$ds^2 = -\varepsilon_1 dz^2 + (-\varepsilon_2)\left(dx^2 + dy^2\right) \qquad (28)$$

(since according to eq.(19) $\varepsilon_l$ is positive and almost constant, as plotted in Fig.3(b)). The scale factor of the effective Minkowski spacetime calculated using eq.(18) is plotted in Fig.2 as a function of $kT/\mu H$ at different values of $-\alpha_\infty \varepsilon_m$. Let us assume that the temperature distribution inside the ferrofluid may be described as

$$T = T_c - z\nabla T = T_c - \eta z \ , \qquad (29)$$

where $T_c$ is the temperature of the metric signature transition (the $\varepsilon_2$=0 point). Since z coordinate plays the role of time in the effective spacetime described by eq.(28), such a temperature gradient $\eta$ will result in a Big Bang-like spacetime expansion described by the scale factor $-\varepsilon_2(T_c - \eta\ z)$ plotted in Fig.3(a). Indeed, as the ferrofluid temperature falls away from the $T_c$ boundary at z=0, the spacetime scale factor increases sharply as a function of z. Note that at larger values of $-\alpha_\infty \varepsilon_m$ expansion of the effective spacetime accelerates at lower temperatures. As demonstrated below, this "cosmological" spacetime expansion may be visualized directly using an optical microscope.

We should also note that in the case of non-constant z-dependent $\varepsilon_l = \varepsilon_x = \varepsilon_y$ and $\varepsilon_2 = \varepsilon_z$ the electromagnetic field separation into the ordinary and the extraordinary components remains well defined [20,21]. Taking into account z derivatives of $\varepsilon_l$ and $\varepsilon_2$, eq.(5) becomes

$$-\frac{\partial^2 \varphi_\omega}{\varepsilon_1 \partial z^2} + \frac{1}{(-\varepsilon_2)}\left(\frac{\partial^2 \varphi_\omega}{\partial x^2} + \frac{\partial^2 \varphi_\omega}{\partial y^2}\right) + \left(\frac{1}{\varepsilon_1^2}\left(\frac{\partial \varepsilon_1}{\partial z}\right) - \frac{2}{\varepsilon_1 \varepsilon_2}\left(\frac{\partial \varepsilon_2}{\partial z}\right)\right)\left(\frac{\partial \varphi_\omega}{\partial z}\right) +$$

$$+ \frac{\varphi_\omega}{\varepsilon_1 \varepsilon_2}\left(\frac{1}{\varepsilon_1}\left(\frac{\partial \varepsilon_1}{\partial z}\right)\left(\frac{\partial \varepsilon_2}{\partial z}\right) - \left(\frac{\partial^2 \varepsilon_2}{\partial z^2}\right)\right) = \frac{\omega_0^2}{c^2}\varphi_\omega \qquad (30)$$

Since $\varepsilon_1$ remains almost constant in a very broad temperature range far from $T=0$ (as evident from Fig.3(b)) its derivatives may be neglected. If we also neglect the second derivative of $\varepsilon_2$, the wave equation for the extraordinary field $\varphi_\omega = E_z$ may be re-written as

$$-\frac{\partial^2 \varphi_\omega}{\varepsilon_1 \partial z^2} + \frac{1}{(-\varepsilon_2)}\left(\frac{\partial^2 \varphi_\omega}{\partial x^2} + \frac{\partial^2 \varphi_\omega}{\partial y^2}\right) - \frac{2}{\varepsilon_1 \varepsilon_2}\left(\frac{\partial \varepsilon_2}{\partial z}\right)\left(\frac{\partial \varphi_\omega}{\partial z}\right) = \frac{\omega_0^2}{c^2}\varphi_\omega \qquad (31)$$

It is easy to verify that the latter equation coincides with the Klein-Gordon equation [22] for a massive particle in a gravitational field described by the metric coefficients $g_{00}=-\varepsilon_1$ and $g_{11}=g_{22}=-\varepsilon_2$:

$$\frac{1}{\sqrt{-g}}\frac{\partial}{\partial x^i}\left(g^{ik}\sqrt{-g}\frac{\partial \psi}{\partial x^k}\right) = \frac{m^2 c^2}{\hbar^2}\psi \qquad (32)$$

Indeed, for a wave field $\psi = \left(-\varepsilon_2\right)^{1/2}\phi$ eq.(31) is reproduced if we also neglect the second derivative terms.

As demonstrated by Fig.1(c), the alignment of cobalt nanoparticles along the timelike z direction is easy to visualize. Moreover, as demonstrated in [18], visibility of the filaments in such images, which may be assessed quantitatively by Fourier analysis, provides measurements of the relative value of $\alpha(H,T)$ compared to $\alpha_\infty$. A natural way to study ferrofluid behaviour near the metric signature transition is to gradually reduce the external magnetic field $H$ until the periodic alignment of cobalt nanoparticles begins to disappear. Since thermal and/or magnetic field gradients are unavoidable in such an experiment, a ferrofluid region showing gradual disappearance of filament periodicity across the image is relatively easy to find. A microscopic transmission image of such a region taken using illumination with λ=1.5 μm light is presented in Fig.4(b). The filament periodicity (compare Fig.4(a) and 4(b)) gradually disappears towards the top of the image, which is verified by Fourier analysis in Fig.4(c) (note that the contrast in these Fourier images is internal). The measured dependence of the spacetime scale factor $-\varepsilon_2$ on the effective time calculated at λ=1.4 μm is shown in Fig.4(d). This wavelength is close enough to the λ=1.5 μm illumination wavelength used in the experiment, while the experimental data for the electromagnetic properties of cobalt [13] may also be used. Calculated values of $-\varepsilon_2$ are based on the Fourier analysis of smaller regions in Fig.4(b) (shown by the yellow boxes). The measured behaviour of the scale factor in this experiment corresponds to small values of $-\alpha_\infty \varepsilon_m$, which is expected at λ=1.4 μm based on the optical properties of cobalt [13]. As evident from Fig.4(d), the measured data show good agreement with theoretical calculations based on eq.(18).

Unfortunately, the described analogy between the extraordinary light propagation inside the ferrofluid and the dynamics of massive particles in Minkowski spacetime is in no way perfect. The main difficulty comes from the cross-coupling between extraordinary and ordinary light inside the ferrofluid, which may be caused by internal defects and strong gradients of the dielectric permittivity tensor components. Since ordinary light "lives" in Euclidean space, such a cross-coupling breaks the effective Lorentz symmetry (6) of the system. In addition, such a model is necessarily limited to 2+1 spacetime dimensions. Nevertheless, these limitations notwithstanding, it is interesting to trace the appearance of macroscopic "gravity" (even though imperfect) to thermodynamics of well understood microscopic degrees of freedom of the ferrofluid.

In conclusion, we have considered thermodynamics of the effective Minkowski spacetime which may be formed under the influence of external magnetic field in a cobalt ferrofluid. It appears that the extraordinary photons propagating inside the ferrofluid perceive thermal gradients in the ferrofluid as an effective gravitational field, which obeys the Newton law. Unlike quantum mechanical derivation based on the holographic principle reported in [2], the Newton law in a ferrofluid arises in a purely classical Boltzmann system. The effective Minkowski spacetime behaviour near the metric signature transition may mimic various cosmological Big Bang scenarios, which may be observed directly using an optical microscope. Even though the considered model is applicable only to the extraordinary photons, it is quite interesting since it permits both direct theoretical analysis at the level of well-defined microscopic degrees of freedom, and it enables direct visualization of these microscopic degrees of freedom using standard optical tools.

.

**Figure Captions**

**Figure 1.** (a) Schematic geometry of the metal nanowire-based hyperbolic metamaterial. (b) Wavelength dependencies of the real parts of $\varepsilon_z$ and $\varepsilon_{x=}\varepsilon_y$ for a cobalt nanoparticle-based ferrofluid at $\alpha=8.2\%$ volume fraction of nanoparticles. While $\varepsilon_{x=}\varepsilon_y$ stays positive and almost constant, $\varepsilon_z$ changes sign to negative around $\lambda=1\mu m$. (c) Microscopic image of cobalt nanoparticle-based ferrofluid in external magnetic field reveals nanoparticle alignment along the field direction.

**Figure 2.** Schematic geometry of the Newton law derivation for effective gravity in a ferrofluid.

**Figure 3**. (a) Scale factor of the effective Minkowski spacetime $g_{11}=g_{22}=-\varepsilon_2$ (calculated using eq.(18)) plotted as a function of $kT/\mu H$ at different values of $-\alpha_\infty \mathcal{E}_m$. It is assumed that $\varepsilon_d=2$. (b) Metric coefficient $g_{00}=-\varepsilon_1$ of the effective Minkowski spacetime calculated using eq.(19).

**Figure 4**. (a) Microscopic transmission image of the cobalt nanoparticle-based ferrofluid taken in external magnetic field $H$ using illumination with $\lambda=1.5$ $\mu$m light. (b) Analysis of gradual melting of the effective Minkowski spacetime within a single microscopic image of the ferrofluid. This image was obtained while the magnitude of external magnetic field $H$ was decreased gradually. The effective time direction is indicated by the red arrow. The effective spacetime expansion is shown schematically by the yellow cone. (d) Measured dependence of the spacetime scale factor $-\varepsilon_2$ on the effective time calculated at $\lambda=1.4$ $\mu$m. These calculations are based on the Fourier analysis (c) of smaller regions of Fig.4(b) shown by the yellow boxes. The measured data are compared with theoretical calculations based on eq.(18).

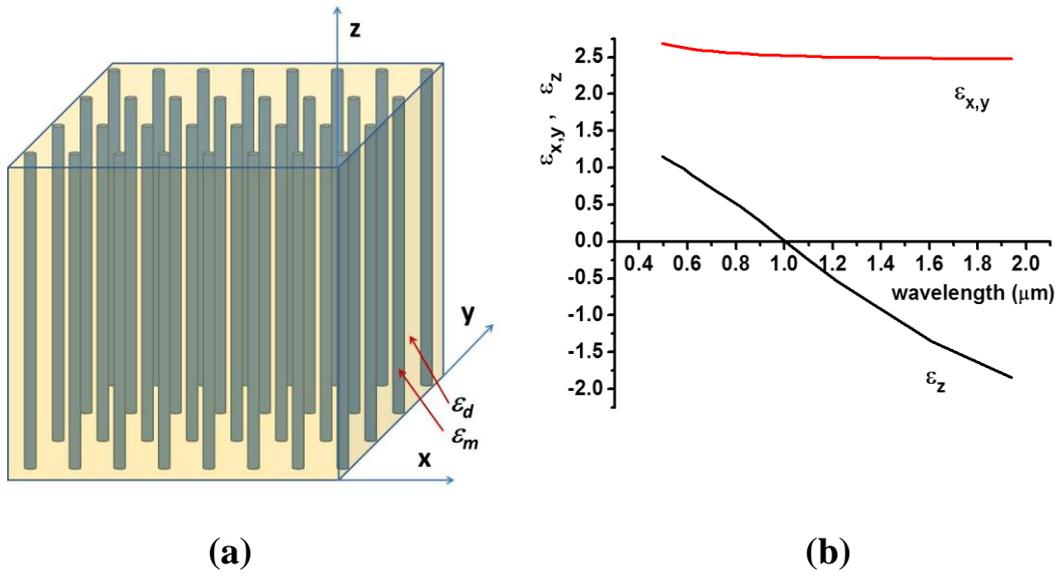

**(a)**                    **(b)**

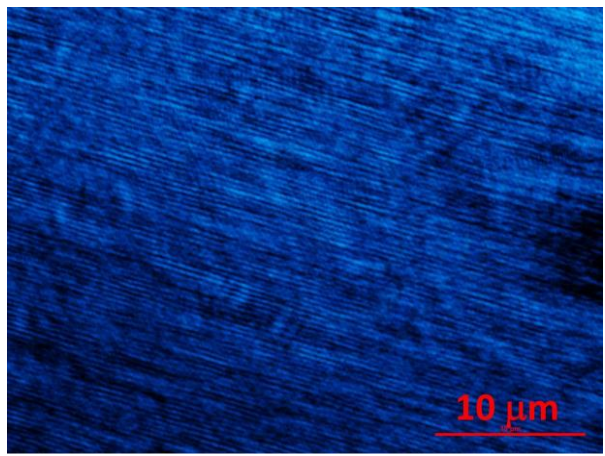

**(c)**

Fig. 1

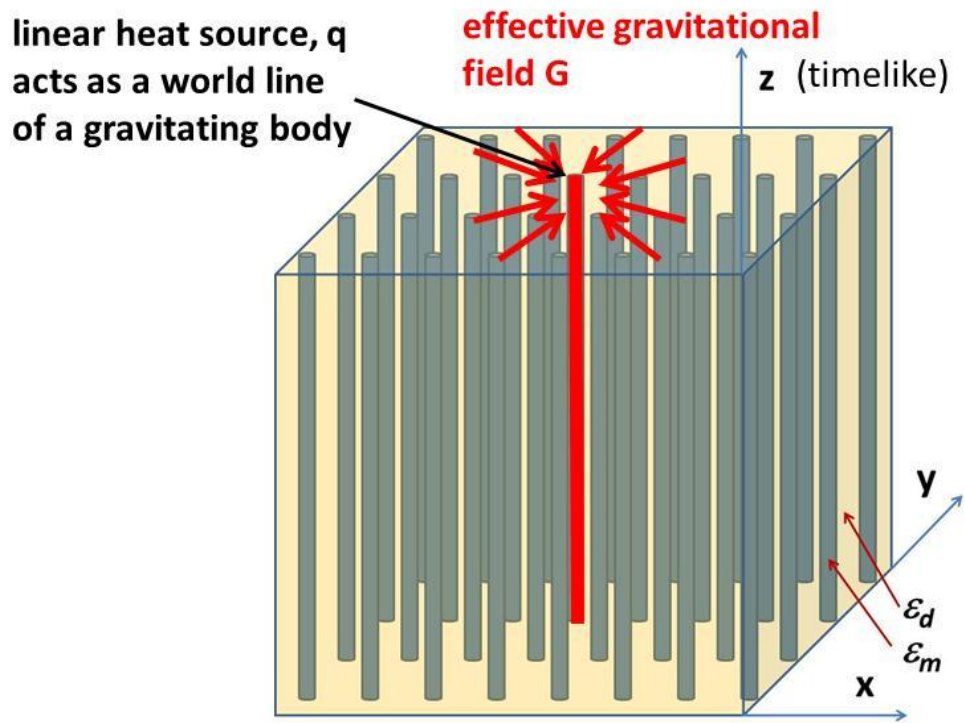

Fig. 2

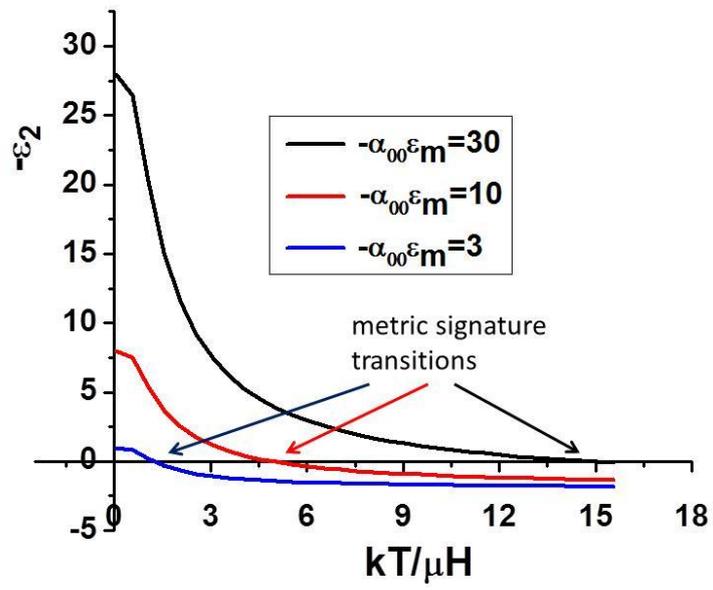

(a)

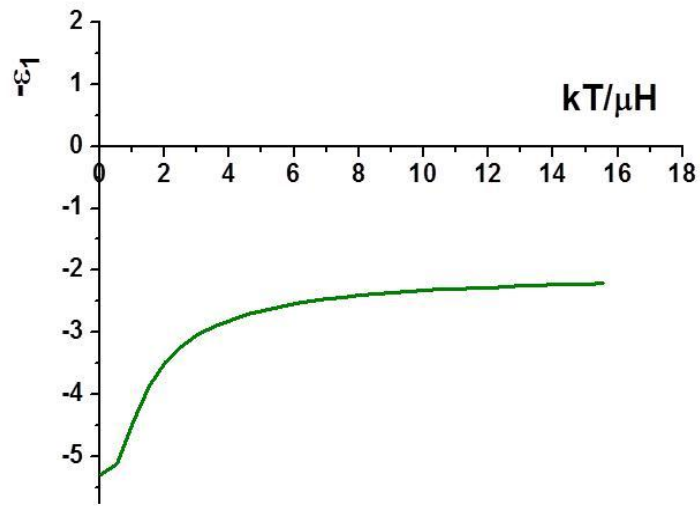

(b)

Fig. 3

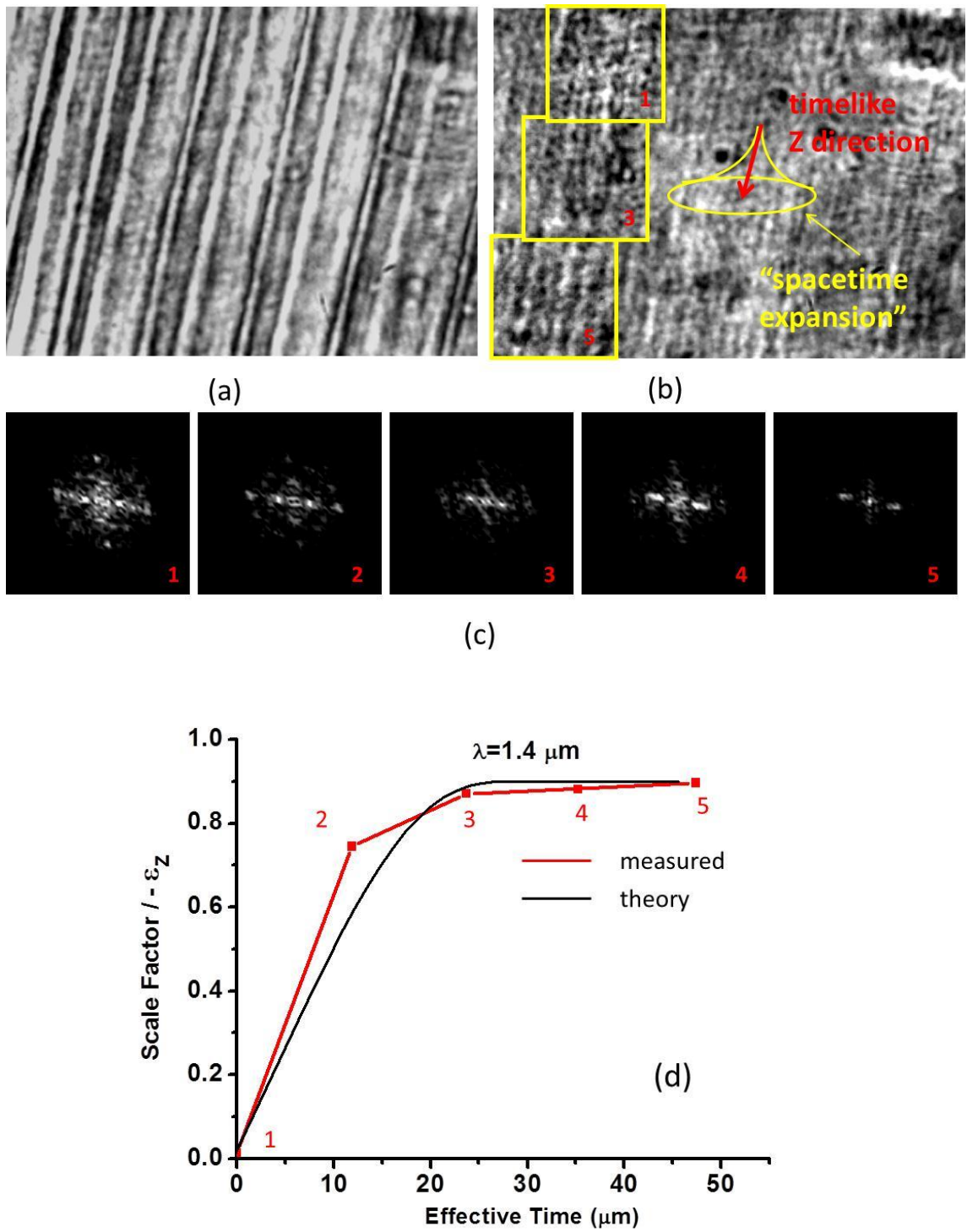

(a)                                          (b)

(c)

(d)

Fig. 4